\documentclass[aps,prl,twocolumn,showpacs,superscriptaddress,amsmath,amssymb]{revtex4}

\usepackage{graphicx}

\begin{document}

\title{Cathodoluminescence imaging and spectroscopy on a single multiwall boron nitride nanotube}

\author{P. Jaffrennou}
\affiliation{Laboratoire d'Etude des Microstructures,
ONERA-CNRS, BP 72, 92322 Ch\^atillon Cedex, France}
\affiliation{D\'epartement de Mesures Physiques, ONERA,
Chemin de la Huni\`ere, 91761 Palaiseau Cedex, France}
\affiliation{Laboratoire de Photonique Quantique et
Mol\'eculaire, Institut d'Alembert, Ecole Normale Sup\'erieure de
Cachan, 61 avenue du Pr\'esident Wilson 94235 Cachan Cedex,
France}
\author{F. Donatini}
\affiliation{Laboratoire de Spectrom\'etrie Physique,
Universit\'e Joseph Fourier, Grenoble I, BP 87, F-38402 Saint
Martin d'H\`eres Cedex, France}
\author{J. Barjon}
\affiliation{Groupe d'Etudes de la Mati\`ere Condens\'ee,
Universit\'e de Versailles St Quentin, CNRS Bellevue, 1 place
Aristide Briand, 92195 Meudon Cedex, France}
\author{J.-S. Lauret}
\affiliation{Laboratoire de Photonique Quantique et
Mol\'eculaire, Institut d'Alembert, Ecole Normale Sup\'erieure de
Cachan, 61 avenue du Pr\'esident Wilson 94235 Cachan Cedex,
France}
\email{lauret@lpqm.ens-cachan.fr}
\author{A. Maguer}
\affiliation{Laboratoire d'Etude des Microstructures,
ONERA-CNRS, BP 72, 92322 Ch\^atillon Cedex, France}
\affiliation{D\'epartement de Mesures Physiques, ONERA,
Chemin de la Huni\`ere, 91761 Palaiseau Cedex, France}
\affiliation{CEA Saclay DSV/DBJC/SMMCB, B\^{a}timent 547, 91191 Gif sur Yvette Cedex, France}
\author{B. Attal-Tretout}
\affiliation{D\'epartement de Mesures Physiques, ONERA,
Chemin de la Huni\`ere, 91761 Palaiseau Cedex, France}
\author{F. Ducastelle}
\affiliation{Laboratoire d'Etude des Microstructures,
ONERA-CNRS, BP 72, 92322 Ch\^atillon Cedex, France}
\author{A. Loiseau}
\affiliation{Laboratoire d'Etude des Microstructures,
ONERA-CNRS, BP 72, 92322 Ch\^atillon Cedex, France}

\begin{abstract}
Cathodoluminescence imaging and spectroscopy experiments on a single bamboo-like boron nitride nanotube are reported. Imaging experiments show that the luminescence is located all along the
nanotube. Spectroscopy experiments point out the important role of dimensionality in this one dimensional object.
\end{abstract}


\maketitle

Since the early 90's, theoretical and experimental studies on nanotubes have attracted much attention. Carbon nanotubes, which have been first discovered in 1991 \cite{Iijima91}, are the most commonly studied. They are investigated in various domains such as mechanics, molecular electronics, optoelectronics and optics. Few years after the discovery of carbon nanotubes, boron nitride nanotubes have also been synthesized \cite{Chopra95}. These nanotubes are either composed of a few (multiwall) or of a single (single wall) rolled-up hexagonal boron nitride sheet(s). Single wall nanotubes are the best systems from a fundamental point of view, but it is often easier to produce multiwall nanotubes. The study of the properties of multiwall nanotubes is then a first step since it can make the link with hexagonal bulk material. BN nanotubes are wide band gap semiconductors with optical transitions in the UV range above 5 eV 
\cite{Rubio94,Blase94,Arenal05,Lauret05,Wirtz06,Park06}. This property makes them very different from carbon nanotubes and attractive for other kinds of applications as blue light and UV emitters. 

Moreover, because of their high thermal stability, BN materials are studied and already used in vacuum technology. Some experiments performed on BN nanotubes have been reported such as structural investigation \cite{Arenal05,Arenal06,Zobelli06,Arenalbis06}, transport \cite{Radosavljevic03,Chang06,Golberg06}, scanning tunneling microscopy and spectroscopy \cite{Czerw03,Ishigami05,Ishigami06}, optical absorption spectroscopy \cite{Lauret05}, cathodoluminescence and photoluminescence spectroscopies \cite{Wu04,Zhi05,Chen06}. The investigation of their optical properties is a key point in the understanding of the fundamental processes and the future applications of these one dimensional (1D) materials. However, their optical properties remain poorly known especially because of their wide band gap.

\section{Experimental details}
\label{Exp}

\begin{figure}
    \centering
    \includegraphics[scale=.85]{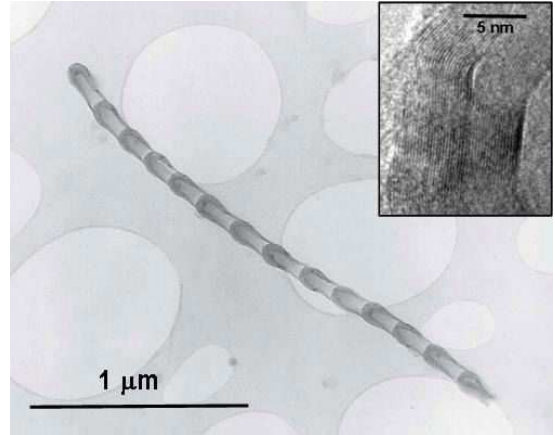}
    \caption{TEM image of a bamboo-like MWBNNT. Inset: zoom on the
nanotube walls.}
    \label{fig1}
\end{figure}

When studying nano-objects, one important step is to perform experiments on an individual object since then, we are free from size fluctuations and other averaging effects which scramble the
intrinsic properties. Until now, only few experiments have been reported concerning electronic properties of a single BN nanotube (Electron Energy Loss Spectroscopy performed in the low loss
regime \cite{Arenal05} and Scanning tunneling Spectroscopy \cite{Ishigami05}). In this Letter, we report the observation of the luminescence of a single BN nanotube by studying
cathodoluminescence imaging and spectroscopy on an individual bamboo-like Multiwall Boron Nitride Nanotube (MWBNNT). MWBNNT studied here are synthesized by ball milling in the group of Pr Ying Chen in the Australian National University \cite{Chen99}. Transmission Electron Microscopy (TEM) analysis demonstrates that the product of the synthesis is mainly composed of bamboo-like structured MWBNNT (Fig.\ \ref{fig1}).

These bamboo-like MWBNNT are composed of about 35 h-BN sheets and their average external diameter is about 50 nm. Typically, there are 30 stacks in the bamboo structure and high resolution TEM images (inset of Fig.\ \ref{fig1}) show the good crystallinity of this bamboo-like MWBNNT. Cracks of the lattice are also observed into each stack, as reported in BN nanorods \cite{Zhang06}.

The MWBNNT are dispersed in ethanol, centrifugated at 25000 g and then deposited on a C-coated Cu TEM grid. Bamboo-like MWBNNT are suspended on the amorphous carbon membrane. The sample is characterized by High Resolution Transmission Electron Microscopy
(HRTEM) using a Philips CM-20 (200 keV) equipped with a LaB$_{6}$ filament. Cathodoluminescence experiments are performed using a 20 keV, 400 pA electron beam in a scanning electron microscope (FEI Quanta 200). The light is collected thanks to a parabolic
mirror and coupled to HR460 (Jobin Yvon) spectrograph equipped with a nitrogen cooled CCD array for spectroscopy and a photomultiplier tube for imaging. Experiments are performed at 5ÊK.

\begin{figure}
    \centering
    \includegraphics[scale=.6]{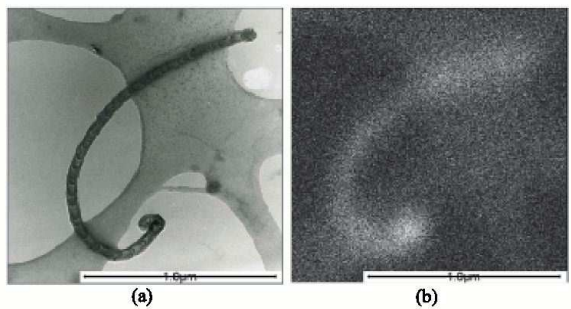}
    \caption{(a)TEM image and (b) Cathodoluminescence image (bottom) at
T=5 K, I=400 pA, of the same individual MWBNNT.}
    \label{fig2}
\end{figure}

The cathodoluminescence image of an individual MWBNNT is displayed in Fig.\ \ref{fig2}b. The comparison between the TEM (Fig.\ \ref{fig2}a) and the CL (Fig.\ \ref{fig2}b) images shows that the emitted light is located all along the nanotube.

\begin{figure}
    \centering
    \includegraphics[scale=.75]{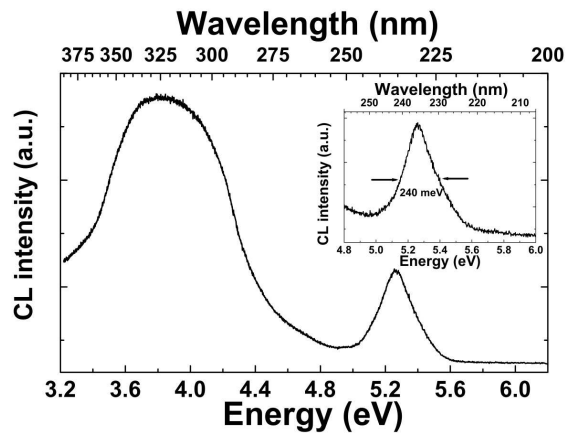}
    \caption{Cathodoluminescence spectrum of the
individual MWBNNT at T=5 K and I=400 pA. Exposition time: 25 min.
Inset: zoom on the UV line.}
    \label{fig3}
\end{figure}

The spectrum of the light emitted by the single MWBNNT is displayed in Fig.\ \ref{fig3}  and exhibits two major lines. A first line is centered at 5.27 eV with a full width at half maximum (FWHM) of 240ÊmeV. A second line is centered on 3.8 eV with a FWHM of about 1 eV. Since excitonic effects in BN materials are expected to govern their optical properties \cite{Arnaud06,Wirtz06,Park06}, the line at 5.27 eV can be attributed to the transitions of the quasi-Frenkel excitons of the bamboo-like nanotube. It has to be compared with the quasi Frenkel excitonic line at 5.77 eV observed in the h-BN photoluminescence and cathodoluminescence spectra \cite{Watanabe04,Watanabe06,Watanabebis06,Silly06}. The red shift of 490 meV can be considered as the spectral feature of the tubular geometry. By rolling up h-BN sheets, some opposite effects can occur inducing blue or red shift of the line: an increase of the band gap energy due to confinement (blue shift) and of the binding energy of the exciton (red shift). The Bohr radius of this quasi-Frenkel exciton can been estimated at about 1 nm and has to be compared with the diameter ($\sim 50$ nm) of the bamboo nanotube. Then, the exciton should not be too much confined by enrolling the sheets. Other effects have to be investigated in this ionic material as internal electric field effects \cite{Khoo04} or trapping of excitons as observed in molecular crystals.

The inset in Fig.\ \ref{fig3} exhibits a zoom on the 5.27 eV line. The linewidth (240 meV at 5 K) is quite large for an individual object as compared to more classical semiconductor heterostructures in which linewidths of Wannier excitons are close to a few meV at 5 K. CL experiments do not show any broadening of the linewidth as a function of current densities. Therefore, many body effects as exciton-exciton interactions should not account for this linewidth. Moreover, the presence of an internal electric field in hexagonal nitrides is known to lead to huge broadening of the lines. Finally, molecular crystals, in which Frenkel excitons are observed, exhibits broad lines. The observation of the 240 meV linewidth of this individual nanotube supports the idea that BN materials seem to behave like molecular crystals.

The line centered at 3.8 eV is attributed to deep levels in the band gap due to intrinsic impurities or defects. The same interpretation is now commonly admitted for the blue line in h-BN \cite{Silly06} and the same band has also been observed in BN nanorods \cite{Koi05,Zhang06}. Furthermore, in previous works on h-BN \cite{Silly06}, noticeable phonon replicas have been
observed in this band with related phonon energy of 0.18 eV. At this energy no phonon mode has been reported in bulk h-BN and, then, the nature of this phonon remains an open question. The
bamboo-like MWBNNT spectrum does not exhibit any sign of phonon replica as observed (in CL and PL) in the 3.8 eV band of BN nanorods \cite{Koi05,Zhang06}. HRTEM images of BN nanrods
\cite{Zhangbis06} and of bamboo-like nanotubes show cracks on the walls. This kind of structure may prevent coupling between electrons and delocalized phonons. This observation support the
interpretation of the 0.18 eV phonon observed in h-BN as a delocalized phonon.

In conclusion, we have reported cathodoluminescence imaging and spectroscopy on an individual bamboo-like multiwall boron nitride nanotube. The luminescence signal is observed all along the nanotube. The effect of the tubular geometry has been observed on the quasi-Frenkel exciton of the bulk h-BN material. As suggested in previous works on h-BN \cite{Watanabe04,Arnaud06,Silly06}, excitonic effects seem to play a major role in the optical response of BN nanotubes, but internal field effects and trapping of excitons have to be
investigated. In order to obtain more intrinsic information about the fundamental processes involved in these wide band gap 1D semiconductors, luminescence experiments on single wall BN nanotubes are in progress.

\section*{Acknowledgements}

The authors are grateful to Pr. Y.
Chen for kindly providing nanotubes sample, and to H. Mariette, L. Wirtz and B. Arnaud for helpful discussions. LEM is a ``Unit\'e mixte" ONERA-CNRS (UMR104). LPQM is a ``Unit\'e mixte" de
recherche associ\'ee au CNRS (UMR8537). GEMAC is a ``Unit\'e mixte" de recherche associ\'ee au CNRS (UMR8635). This work has been supported by the GDR-E ``nanotube" (GDRE2756).

\end{document}